\documentclass[a4paper,fleqn]{cas-sc}
\ExplSyntaxOn
\cs_gset:Npn \__first_footerline:
  { \group_begin: \footnotesize \group_end: }
\cs_gset:Npn \__second_footerline:
  { \group_begin: \footnotesize \group_end: }
\ExplSyntaxOff
\usepackage[numbers,square,sort&compress]{natbib}
\usepackage{comment}
\def\tsc#1{\csdef{#1}{\textsc{\lowercase{#1}}\xspace}}
\tsc{WGM}
\tsc{QE}
\tsc{EP}
\tsc{PMS}
\tsc{BEC}
\tsc{DE}

\begin{document}
\let\WriteBookmarks\relax
\def\floatpagepagefraction{1}
\def\textpagefraction{.001}
\shorttitle{Towards Self-Organizing Channel Coding in 6G-Enabled IoT Networks}
\shortauthors{M. Moazam Azeem et~al.}


\title[mode = title]{Towards Self-Organizing Channel Coding in 6G-Enabled IoT Networks: Trends, Cognitive Architectures, and Open Challenges}                      



                         
\author[1]{M. Moazam Azeem}


\affiliation[1]{organization={Department of Computer Science and Engineering, Qatar University},
                city={Doha},
                country={Qatar}}
\author[2]{M. Adil Raja}
\author[3]{Ayaz Ahmad}

\fntext[fn1]{ORCID: 0000-0002-1644-7240}


\affiliation[2]{organization={Centre for Robotics and Intelligent Systems (CRIS), University of Limerick},
                country={Ireland}}
\affiliation[3]{organization={Department of Electrical Engineering, King Faisal University},
                city={Hofuf},
                country={KSA}}
\cortext[cor1]{Corresponding author: Muhammad Moazam Azeem: email: moazam@qu.edu.qa}


\begin{abstract}
The sixth-generation (6G) of wireless communication is designed to facilitate 
distributed edge intelligence, ultra-reliable low-latency communication (URLLC), 
and massive machine-type communications (mMTC), all of which have the potential 
to revolutionize the Internet of Things (IoT). Autonomous vehicles, swarm robotics, 
XR, and industrial automation are IoT-centric 6G applications with demanding 
requirements; developing efficient, adaptive, and reliable channel coding methods 
is crucial to realizing this goal. This study explores the adaptation of various 
channel coding schemes to distributed IoT scenarios in 6G networks, covering both 
established and developing schemes, including LDPC, Turbo, Polar, and Fountain 
codes. We evaluate the relevance of recent developments to decentralized IoT 
designs, including AI-assisted cognitive coding, neural network-based decoding, 
adaptive and context-aware coding, quantum and reconfigurable intelligent 
surface (RIS) coding. We further propose CACADe-6G, a cognitive architecture for 
adaptive channel coding; our evaluation shows that its principal benefit over 
simpler adaptive baselines lies in reduced switching overhead and energy 
consumption, rather than a guaranteed reduction in bit error rate relative to a 
memoryless threshold comparator. We compare different channel coding schemes with 
respect to latency, energy, and reliability requirements, and highlight important 
use cases. Findings from this study reveal current research gaps and identify 
potential future directions for coding in intelligent, large-scale IoT systems, 
aiming to guide future research and support the deployment of 6G-enabled IoT 
technologies.
\end{abstract}



\maketitle

\section{Introduction}
The emergence of 6G wireless networks promises a communication infrastructure that is hyper-connected, intelligent, and widespread by the end of the 2030s, representing a significant advancement beyond the capabilities of 5G. The 5th generation (5G) delivers enhanced data throughput, extensive connectivity, and reduced latency. Nonetheless, 6G significantly elevates capabilities by facilitating mission-critical applications such as the Tactile Internet, holographic telepresence, intelligent transportation, and the seamless integration of terrestrial, aerial, and satellite systems. 6G networks are designed to include a more distributed and decentralized architecture, facilitating enhanced scalability and improved support for intelligent and collaborative decision-making among devices \cite{ref1_6G_vision}. 6G must facilitate an extraordinary quantity of interconnected devices within the Internet of Things (IoT). These devices may range from battery-operated sensors to autonomous mobile nodes. These IoT systems require connectivity which is not just reliable and rapid but also energy-efficient and effective in resource-constrained environments. 
In such settings, providing reliable and effective communication across possible disruptive and unreliable channels is more challenging. Channel coding, which incorporates additional bits into data streams to detect and rectify errors, is crucial for achieving the ambitious objectives of 6G. Conventional coding schemes, although vital in legacy systems, struggle to fulfill the demands of ultra-low latency (e.g., 100 µs), ultra-high reliability (e.g., 99.99999\%), and the extraordinary scale of devices (e.g., billions of IoT nodes) \cite{ref2_8705373}. Moreover, due to the dynamic nature of distributed networks, there are stringent constraints on coding schemes resulting from variations in topology, mobility, interference, and resource availability. In this IoT-centric environment, channel coding must evolve to accommodate the requirements of low-power, distributed devices communicating via frequently dynamic and lossy connections. 
Due to these tendencies, it is imperative to re-evaluate and innovate channel coding schemes that are not only efficient and robust but also adaptable, intelligent, and resource-conscious \cite{ref3_8594709}. Innovative techniques, including Low-Density Parity-Check (LDPC) codes, Polar codes, Digital Fountain/Rateless Fountain codes, and AI-enhanced encoding/decoding systems, have drawn significant attention. The integration of channel coding with cutting-edge technologies, like reconfigurable intelligent surfaces (RIS), quantum communication, and cell-free massive MIMO, generates tremendous research opportunities \cite{ref4_RIS_MIMO_coding}.

Within the framework of decentralized 6G networks, this research delves deeply into the topic of channel coding. Channel codes for 6G's distributed framework must be nimble across diverse links, scalable to enormous device populations, and optimized for both centralized and edge processing scenarios, in contrast to typical monolithic systems. This paper has five main objectives: (i) to provide a comprehensive overview 
of the various channel coding schemes applicable to 6G networks, both 
classical and emerging; (ii) to evaluate whether these schemes are suitable 
for addressing the requirements of distributed networks in terms of latency, 
reliability, and energy efficiency; (iii) to highlight recent advancements, 
including low-complexity decoders, coding for intelligent environments, and 
machine learning-assisted coding; (iv) to propose and evaluate CACADe-6G, a 
cognitive architecture for adaptive channel coding that reduces switching 
overhead and energy consumption relative to simpler adaptive baselines under 
dynamic channel conditions; and (v) to identify open research challenges and 
propose directions for future work toward standardizing next-generation 
coding frameworks. Wireless communication researchers, systems architects working on 6G infrastructure, and legislators participating in standardization organizations such as 3GPP, ITU, and IEEE are all potential readers of this article.

Following is a summary of the most important findings from this study: Conventional schemes, modern capacity-approaching designs, and newer techniques that reach beyond existing architectures are the three areas into which existing channel coding schemes are organized. We present a comprehensive analysis of well-known coding methods, including LDPC, Turbo, Polar, and Fountain codes, comparing them with respect to energy consumption, decoding complexity, and block error rate (BLER). AI-assisted encoding/decoding, coding for RIS-enabled channels, and the influence of distributed storage and edge intelligence on coding techniques are some of the topics in cross-layer and intelligent design that we investigate. Real-time adaptation, privacy-aware code design, and hybrid coding models are some of the important technical challenges that we have identified. Real-time, context-aware coding in distributed 6G networks is supported by our proposed Cognitive Architecture for Adaptive Channel Coding (CACADe-6G), which blends sensing, learning, and action layers to reduce 
switching overhead and energy consumption under dynamic channel 
conditions. For Internet of Things (IoT) devices, these metrics are essential for achieving a balance between performance and limitations such as short battery life, lightweight computing power, and dense device coexistence. Communication module needs to be cognitively adaptive, self-optimizing, and reliable in dynamic environments.

The rest of this paper is organized as follows. Section~2 describes the background and the main 
requirements of 6G-enabled IoT distributed networks. Section~3 discusses modern and AI-assisted 
channel coding techniques for 6G environments. Section~4 discusses the main challenges of cognitive 
channel coding and presents the proposed CACADe-6G architecture, including its performance 
evaluation. Section~5 presents a comparison of various coding techniques and their application 
scenarios. Section~6 discusses future research directions for 
6G-enabled IoT networks, and Section~7 concludes the paper.

\section{Background and Requirements of 6G-Enabled IoT Distributed Networks}
The emerging 6G architecture is expected to support a vast range of IoT-driven services requiring ultra-low latency, high reliability, and massive connectivity. Distributed 6G networks will play a key role in enabling smart environments such as cities, factories, and healthcare systems by interconnecting billions of autonomous devices operating in heterogeneous and dynamic conditions. These networks must be optimized for the stringent requirements of IoT systems, particularly in terms of scalability, energy efficiency, and decentralized processing.

\subsection{Characteristics and Requirements of 6G}
Sixth-generation (6G) networks are designed to extend beyond the capabilities of 5G by targeting extreme performance metrics: sub-millisecond latency, up to 1 Tbps peak data rates, 99.99999\% reliability, and support for more than 10 million devices per square kilometer. Section~\ref{sec:cacade_eval} evaluates adaptive code selection under a deep-fade
regime chosen to stress-test switching behaviour rather than to
demonstrate this reliability figure directly; Section~5.2 shows the
corresponding high-SNR error-floor behaviour relevant to URLLC
compliance. To achieve these goals, 6G will incorporate intelligence, adaptability, and deep integration between communication, computing, and sensing \cite{ref5_6G_intelligence,ref6_6G_survey_2022}.
Key characteristics of 6G include:
\begin{itemize}
    \item \textbf{Pervasive Intelligence} Embedding machine learning (ML) at the network edge and core for autonomous decision making and optimization.%
    \item \textbf{Native AI Integration} AI-driven resource allocation, routing, and code adaptation become fundamental parts of the stack.%
    \item \textbf{Distributed and Heterogeneous Architectures} Deployment of decentralized units such as edge nodes, small cells, UAVs, and satellites.%
    \item \textbf{Ultra-Dense Networks} High device density and multiple simultaneous connections require scalable protocols and robust error correction.%
    \item \textbf{Energy-Aware Operation} Efficient communication and computation are essential, especially for battery-constrained IoT nodes.%

\end{itemize}
These stringent requirements necessitate rethinking the design of all protocol layers, especially physical and link layers where channel coding resides, to ensure end-to-end reliability and efficiency.

\subsection{Key 6G Use Cases}
Several scenarios are shown below to illustrate why improvements to 6G channel coding are necessary: Channel coding approaches employed in smart agriculture, industrial IoT, and environmental monitoring need to be durable yet lightweight to accommodate the huge number of regularly linked, low-power sensors utilized in these applications. Deterministic performance with latency below one millisecond and extremely low error rates is required in situations requiring safety-critical factors, such as autonomous vehicles, automated manufacturing, and remote healthcare. This is accomplished by Ultra-Reliable Low-Latency Communication (URLLC). Enhanced mobile broadband (eMBB) enables high-throughput channel codes with reasonable decoding complexity, which are required for high-capacity applications such as 16K video streaming, virtual reality (VR), and holographic communication. ISAC, which stands for "integrated sensing and communication," coding techniques for joint radar communication systems must be flexible enough to accommodate constraints in data transfer and sensing. 
Collaborative learning and inference rely on the interchange of gradient or model changes across network nodes, which necessitates error-resistant and low-latency communication. Here is when AI with distributed processing powers becomes useful. Because of the diverse latency, energy consumption, reliability, and bandwidth needs across these use cases, traditional coding approaches are insufficient \cite{ref7_7529226}.

\subsection{Architecture of Distributed 6G Networks}
\begin{figure}
	\centering
   \includegraphics[width=0.85\textwidth]
    {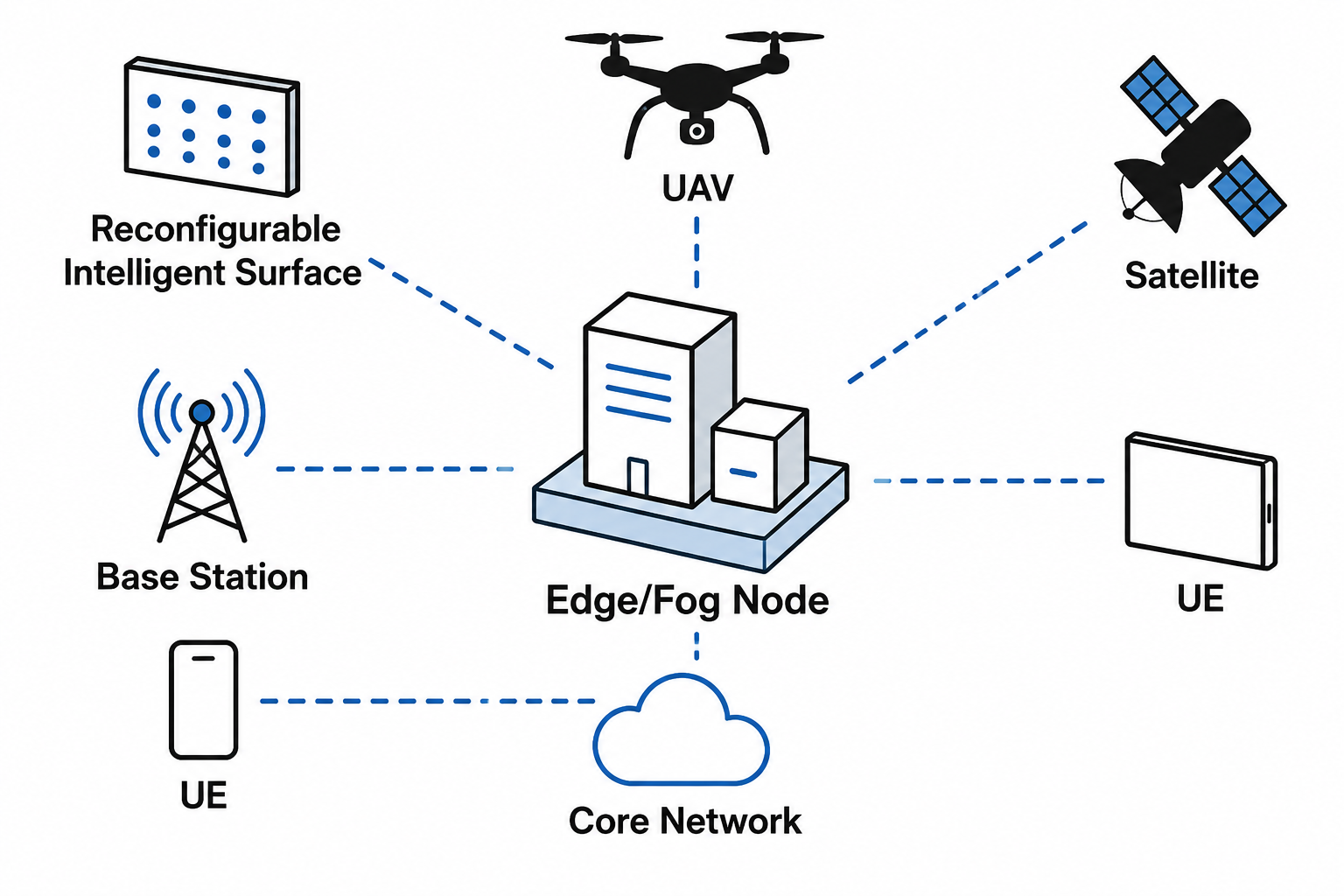}
	\caption{6G Distributed Network Architecture}
	\label{Figure:1}
\end{figure}

Expected 6G features include a distributed design that allows for flexible, scalable, and fault-tolerant communication. Fig.~\ref{Figure:1} shows the main parts of the system. Included in this category are the following nodes: edge/fog, core, RIS, base, UAV, UE, and satellite. Dashed lines are a great way to represent wireless or virtual connections. By relocating processing closer to the data source, edge and fog computing nodes lower latency and relieve load on the core network. For localized processing capabilities to be supported, channel coding needs to advance. Massive MIMO Alternative to Cells: Instead of using individual cells, this technique uses a distributed antenna network to support multiple users at once. Methods of synchronized and interference-aware channel coding are required for this configuration. In order to optimize the propagation of wireless communications, reconfigurable intelligent surfaces (RIS) can be designed to reflect and shape the signals. Channel properties are subject to dynamic changes due to RIS, and the code should be able to adapt to these changes. The requirement for delay-tolerant and rateless codes is brought to light by the fact that the 3D coverage layer uses platforms both on the ground and in the air, creating new channels that are extremely responsive to Doppler and high-delay situations. In the aftermath of a tragedy, these networks are vital, particularly in outlying areas. In order for this design to address issues like distributed interference, frequent handovers, and link heterogeneity, adaptive and intelligent channel coding methods are required \cite{ref8_adaptive_coding_distributed}.

\subsection{Foundations of Channel Coding}
\subsubsection{Basic Principles and Terminology}
The bit rate of a the codeword is defined as $R = k/n$, that is the ratio of information bits to total bits of codeword. If redundancy is increased, the code rate is decreased. The latency and error-correcting capabilities are also affected by the codeword's length n, which is a block length parameter. The Minimum Distance $d_{min}$, also called the Hamming distance, is the minimum number of bits that differ between any two codewords in the code. It determines the maximum number of errors that can be corrected. Some codes only detect errors; others apply techniques such as syndrome decoding or iterative algorithms to fix them. Hard decoding uses single-value decisions (either 0 or 1), whereas soft decoding uses probabilities or confidence levels, typically yielding better results. Revaluating these fundamentals for distributed 6G networks requires new design trade-offs in light of significant latency, energy cost, and codeword length constraints \cite{ref10_coding_fundamentals}.

\subsubsection{Traditional Coding Techniques}
Before capacity-approaching codes were developed, conventional coding techniques were the norm in wireless and satellite networks. Although they may not meet modern performance goals, these methods are suitable for use in some 6G scenarios, including as low-power IoT, due to their conceptual simplicity and hardware efficiency. 
Among the most fundamental forms of conventional codes, block codes have a constant length. These codes convert a k-bit message into an n-bit codeword. Two listings of families with a lot of research are these: Codes using Hamming designed to correct errors that impact a single bit. Because they are brief and to the point, these block codes work well with control channels and older legacy systems. Bose-Chaudhuri-Hocquenghem (BCH) codes generalise Hamming codes to correct multiple random bit errors within a block, trading a modest increase in decoding complexity for substantially stronger error-correcting capability while retaining deterministic, low-latency decoding~\cite{ref80_BCH_edge}. This makes BCH codes attractive for short-block, latency-critical scenarios, as discussed further in Section~5.4.4.. Reed-Solomon (RS) Codes are non-binary codes that have the potential to correct multiple symbol mistakes; they are among the most popular block codes used in storage and broadcast applications such as CD and DVB. 
Because of their computational efficiency for small block sizes, algebraic approaches like syndrome decoding or Euclidean algorithms are often used to decode these block codes. 
Convolutional codes are the second kind of conventional codes. These codes convolve the input sequence with generator polynomials to encode data streams. They differ from block codes in that they can recall previous inputs, which helps with error correction across time-correlated channels. Convolutional codes are often decoded using the Viterbi algorithm, a maximum likelihood decoder, in 2G/3G cellular systems, space missions, and early Wi-Fi protocols. Because it determines the number of input bits that affect each output, the code length also impacts the efficiency of convolutional codes. Longer code lengths make decoding more difficult, but they also increase efficiency. Even if Turbo and LDPC codes are more successful, convolutional codes still provide a good complexity-performance trade-off for low-latency, short-packet communications in environments with limited resources.

\subsubsection{Evolution Toward Modern Codes}
Analytical tractability was a primary goal in the design of traditional codes. To counter this, Shannon's channel capacity theorem established a theoretical upper limit on the maximum data rate that may be safely communicated via a channel that is experiencing noise \cite{ref11_shannon}. Decades of study have been directed towards finding a practical way to overcome this barrier.
\begin{itemize}
    \item \textbf{LDPC Codes} LDPC codes provide near-capacity performance when it comes to iterative belief propagation decoding; they were first developed by Gallager in 1962.%
    \item \textbf{Turbo Codes} Iterative decoding and parallel concatenated convolutional codes were introduced by Turbo Codes.
    \item \textbf{The Fountain (Rateless) Code} This is a kind of rateless erasure code that was developed for efficient data transmission across unreliable or multicast channels. It is similar to the Luby Transform (LT) code and the Raptor code.%
    \item \textbf{Polar Codes} These are the codes that achieve capacity with low complexity in symmetric binary input memoryless channels, including encoding and decoding.%
    \end{itemize}
These contemporary codes provide the backbone of all communication systems, both present and future. Decoding complexity, energy efficiency, latency, and flexibility must be carefully considered for their practical application in 6G distributed networks. In the future of distributed intelligence in IoT networks, channel coding will be an adaptive decision-making component built into cognitive communication agents, rather than a static process.

\section{Modern and AI-Assisted Channel Coding Techniques in 6G}
\subsection{Modern Coding Techniques}
Recent advances in channel coding have made it possible to achieve near-capacity performance with much lower latency and more scalability than was previously possible with conventional codes. Some examples of such codes include Turbo, Low-Density Parity-Check (LDPC), Polar, and Fountain. Strong candidates for 6G implementation in distributed network scenarios, each of which has shown success in different wireless standards and provides unique benefits under different limitations \cite{ref12_9328621}. 
Low-Density Parity-Check linear block codes, often known as LDPC codes, are defined by a sparse parity control matrix. Iterative belief propagation techniques decode them with performance that is near the Shannon capacity. The LDPC codes include a bipartite Tanner graph that is composed of the nodes that represent variables and check nodes. During decoding, the sparsity guarantees computing efficiency. Modern communication technologies such as IEEE 802.11 (Wi-Fi), 5G NR data channels, DVB-S2, and WiMAX employ LDPC codes. Their Hardware implementation, high throughput, and decoding can possibly be parallelized.  However, Short block lengths LDPC lead to worse performance, and their iterative decoder's delay might vary. In 6G distributed networks, LDPC codes are promising for high-data-rate backhaul and edge computing scenarios where long packets and decoding resources are available \cite{ref13_LDPC_6G}.
Using iterative decoding and concurrent concatenation of convolutional codes, turbo codes may get very near to channel capacity, which is particularly useful for blocks of medium durations. Turbo codes are built of many interconnected recursive convolutional encoders. These are used in extensive deep-space applications, satellite communications, and 3G/4G mobile systems. They have excellent performance for codewords of modest length and effectiveness in burst error settings. However, their iterative nature makes decoding highly difficult and latency-intensive; not ideal for applications requiring extremely low latency. 
There are a few 6G use cases where Turbo codes might be useful, such as control plane communication or situations requiring strict adherence to legacy infrastructure compatibility \cite{ref14_turbo_codes}.

The first provably capacity-achieving codes for binary-input memoryless symmetric channels were polar codes, which were developed by Arıkan in 2009. Their design is based on the polarization of channels, which turns them into subchannels that are either very dependable or very unstable. Their low encoding complexity and strong performance at small block lengths led to their use in 5G NR for control channels. They have compatibility with short-packet transmission, have a solid theoretical basis, and have low-complexity successive cancellation List (SCL) decoding. Polar codes can be ideal in 6G for applications requiring short-packet, ultra-reliable communication, such as haptic internet and real-time command and control \cite{ref15_polar_codes_arikan}.
There is a class of codes known as Fountain and Rateless Codes. This class includes codes like LT (Luby Transform) and Raptor codes, which can generate an endless number of encoded symbols until decoding is successful; each output symbol is a random combination of the input symbols. Once you get enough of them, you can decode them. Rateless codes are great for situations where the channel conditions are uncertain or dynamic, such as UAV swarms or satellite networks, as well as for broadcasting and multicasting. These codes require no retransmissions in the event of channel failure; they require little feedback. Reliance on probabilistic performance guarantees and higher decoding complexity compared to standard codes are limitations. In lossy environments, when retransmissions are expensive or not practicable, fountain codes could offer promising service for 6G applications that include broadcast or large-scale content distribution \cite{ref16_fountain_codes}. In addition, IoT devices with limited resources that send out little packets on an intermittent basis across unreliable connections work well with Fountain codes. Because they are not time-dependent, distributed sensor networks can easily accommodate their irregular transmission patterns. Table \ref{tbl1} provides a concise overview of current research on channel coding for 6G distributed networks.
\begin{table}
\caption{SUMMARY OF RECENT WORKS ON CHANNEL CODING FOR 6G AND DISTRIBUTED IOT NETWORKS}
\label{tbl1}
\centering
\begin{tabular}{p{3cm} p{4cm} p{4cm} p{3cm}}
\toprule
Channel	Coding Technique
 & Key Contributions & Application Context & References \\
\midrule

LDPC Codes & Structured sparse codes with iterative belief propagation decoding & High-throughput backhaul, mMTC, edge computing & \cite{ref18_LDPC_edge, ref21_LDPC_backhaul, Ren2024GAMS, Wang2024NeuralLDPC, Kim2024NSCMS, Guo2024BitEmbedding, Li2024OSD} \\

Turbo Codes & Parallel concatenated convolutional codes with iterative decoding & Legacy systems, distributed storage, control channels & \cite{Xia2024Turbo, ALEBADY2023102010} \\

Polar Codes & Capacity-achieving codes with channel polarisation & URLLC, short-packet communications, XR & \cite{8962344, 7055304, 10106132, 10061665} \\

Fountain Codes & Rateless codes for erasure channels & UAV networks, multicast, satellite communications & \cite{ref28_fountain_UAV, ref29_fountain_satellite} \\

AI-Assisted Coding & Neural decoders, adaptive coding, end-to-end learning & Dynamic IoT environments, edge intelligence & \cite{ref30_AI_coding_dynamic, ref31_AI_coding_edge, 10274838, 7926071}\\

Short Block-Length Codes & Finite blocklength optimization for low latency & URLLC, industrial IoT, vehicular networks &  \cite{8486997} \\

Energy-Efficient Coding & Low-complexity decoding for constrained devices & Battery-powered IoT sensors, wearable devices & \cite{Hyla2023RaptorLDPC, 10802915} \\

Quantum-Enhanced Coding & Quantum LDPC codes, entanglement-assisted schemes & Secure backhaul, quantum IoT applications & \cite{9749227, Dehingia2026QuantumIoT} \\

RIS-Aware Coding & Coding optimised for reconfigurable intelligent surfaces & Smart environments, dynamic channel conditions & \cite{ref40_RIS_coding, 8647620, ref42_RIS_adaptive_coding} \\

Hybrid Coding Schemes & Adaptive multi-service coding frameworks & Heterogeneous IoT deployments & \cite{9745429, 7282910, 7247339} \\

\bottomrule
\end{tabular}
\end{table}
Research into rateless and short block-length codes for URLLC and huge IoT applications is on the rise, despite the fact that LDPC and Polar codes are currently the mainstream.

\subsection{AI-Assisted and Cognitive Coding}
In response to the complex requirements of 6G distributed networks, industry leaders are developing novel approaches to channel coding that surpass traditional design methodologies. These developments leverage AI, quantum communication, and the unique characteristics of upcoming 6G architectural components, such as intelligent surfaces that can be reconfigured and networks that do not require cells. In this section, we will go over a few important themes that are shaping channel coding's future. 
\subsubsection{AI-Assisted Channel Coding} The integration of Artificial Intelligence (AI), particularly Machine Learning (ML) and Deep Learning (DL), into the physical layer of wireless communication networks is reshaping channel coding design for 6G distributed systems \cite{10702508}. The conventional channel coding techniques are more efficient with fixed hardware and static channel assumptions in steady environments. However, when operating in dynamic, heterogeneous, and resource-constrained environments, 6G networks must be able to adapt, be aware of their context, and make calculations in real-time. Channel coding with the help of artificial intelligence (AI) could be the solution to these issues. One example of an architecture that incorporates AI-driven adaptation is CACADe-6G, which will be discussed in section IV. This architecture uses sensing, learning, and decision-making to optimize channel coding in real-time based on application requirements and environmental variables. 

\begin{figure}
	\centering
	\includegraphics[width=0.85\textwidth]{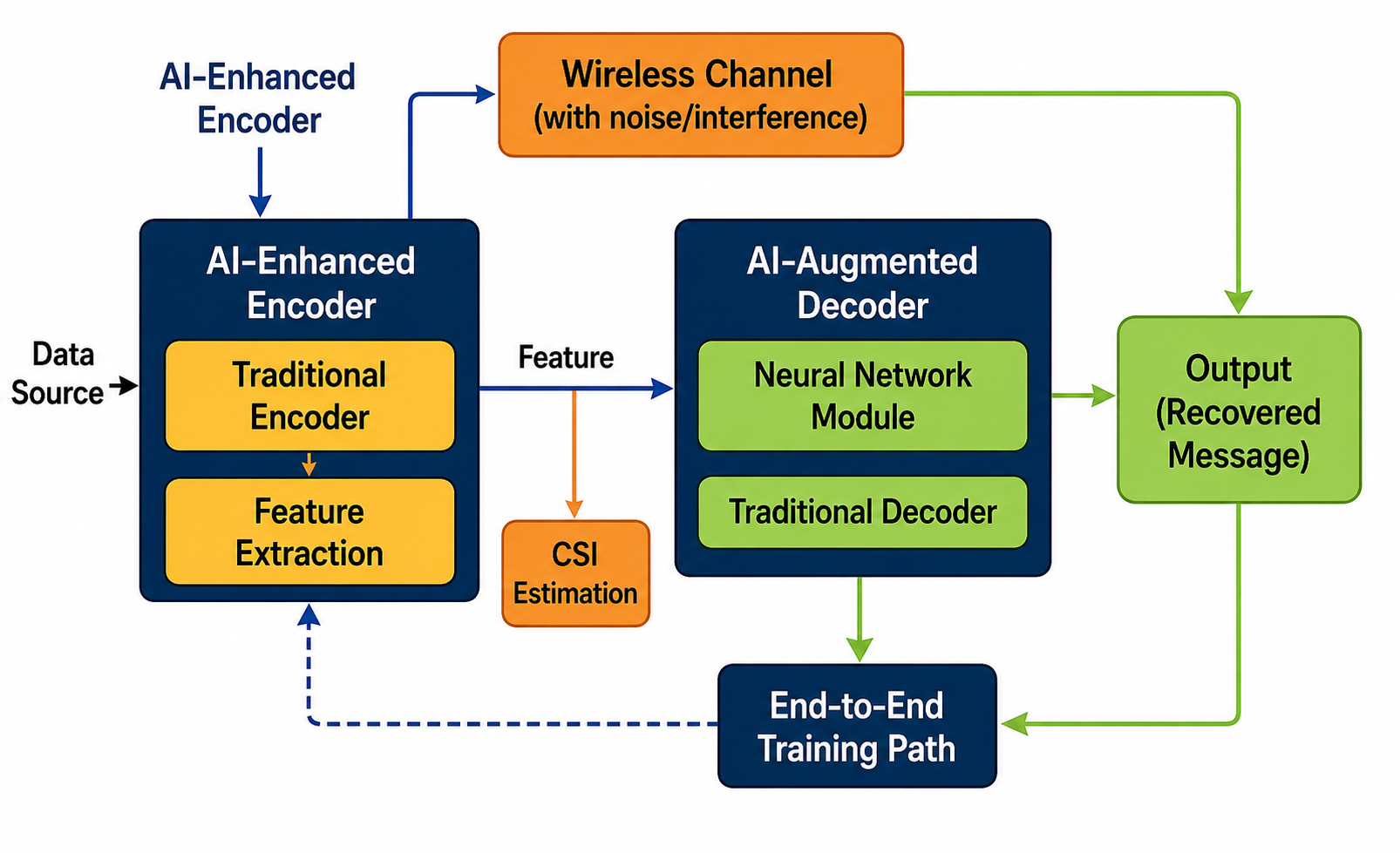}
	\caption{AI-Assisted Channel Coding Framework}
	\label{Figure:2}
\end{figure}

Fig. \ref{Figure:2} shows a broader trend when systems are becoming flexible and data-driven in their approach to adapt to the changing wireless environment. This demonstrates that AI is no longer a supplementary feature but rather a core component of the 6G physical layer. Fig.~\ref{Figure:2} is further explained in the upcoming subsections.                               

\subsubsection{Neural Network-Based Decoders} AI techniques have been successfully applied to replicate or enhance traditional decoding algorithms. For example, deep neural networks (DNNs) have been trained to perform belief propagation for LDPC decoding and successive cancellation List for Polar codes. These neural decoders are often capable of approximating optimal decoding performance with reduced complexity in specific environments \cite{ref47_DNN_LDPC_decoder, ref48_neural_polar_decoder, ref49_neural_decoder_survey}. These Neural Network-Based Decoders have the capacity to learn noise characteristics and are resilient to hardware defects. These codes also have lower latency under specified channel models. During the fast-changing environments, these are perfect for edge devices in intelligent transportation systems or unmanned aerial vehicles.
\subsubsection{End-to-End Learning with Autoencoders} A differentiable autoencoder that treats the whole communication system as a unit from encoder to decoder is another novel approach. This enables joint optimization of all components under realistic channel models, including fading, interference, or non-Gaussian noise \cite{8262721, ref51_e2e_learning, ref52_joint_optimization}. These systems are trained using gradient descent techniques, often requiring simulated data reflecting the target channel conditions. While performance may be high with end-to-end Learning with Autoencoders, understanding what the neural models have learned remains an open challenge.
\subsubsection{Adaptive and Context-Aware Coding} Models trained with artificial intelligence can adjust coding parameters in real time according to known or expected context, such as device mobility, signal-to-noise ratio (SNR), or application delay requirements. These parameters include block length, rate, and decoding iterations. Optimization of hybrid automated repeat requests (HARQs) and dynamic coding rate selection have both been investigated using reinforcement learning \cite{10570511, e27060629, Lin2023AHARQ}. For example, in V2X, the vehicle's speed and channel profile change over time, reinforcement learning agents in vehicular networks can choose the optimal coding scheme.

\subsubsection{Integration in 6G Distributed Architectures} The integration of AI-assisted channel coding enables localized learning and adaptation in distributed topology of 6G, which includes cell-free massive MIMO, RIS, and edge AI platforms. A small artificial intelligence model that adapts encoding and decoding to local users or links can be maintained by any node or cluster of nodes.

\subsection{Emerging Trends (Quantum Computing, RIS, Cell-Free Mesh Architecture)}
\subsubsection{Quantum Channel Coding} In 6G, investigation of ultra-secure and high-capacity networks has piqued interest in quantum communication. In order to put quantum networks into practice, quantum error correction (QEC) algorithms are required \cite{ref55_QEC_networks, ref56_quantum_LDPC, ref58_quantum_6G}. Quantum LDPC Codes are crafted to guarantee scalability and minimal complexity while shielding qubits from quantum noise and decoherence. Using pre-shared entanglement, "entanglement- assisted coding" enhances the efficiency of both classical and quantum error correcting codes. In the future, secure 6G backhaul lines, quantum teleportation technologies, and quantum key distribution (QKD) will all benefit from this. Although QEC is still in its early stages, its concepts are anticipated to impact sophisticated coding techniques in classical systems.
\subsubsection{Code Design for Reconfigurable Intelligent Surfaces} RIS are a game-changing technology that can, via programmable metasurfaces, alter the environment for wireless propagation. Changes in the Effective Channel due to RIS, channel codes need to be able to adjust fast or operate with partial channel state information \cite{ref59_RIS_CSI, ref60_RIS_beamforming, 9362274}. If you want your signals to be as reliable as possible, you should optimize your coding in combination with your RIS beamforming and reflection techniques. Since RIS pieces are often passive, efficient feedback and simple coding are required for practical implementation, which leads to reduced overhead. 
An exciting new field of study, RIS-aware coding has great promise for 6G networks that are both energy efficient and have minimal latency. 
\subsubsection{Coding in Cell-Free and Mesh Architectures} 6G architecture relies heavily on cell-free massive MIMO and wireless mesh networks, which are particularly useful in extremely dense or infrastructure-free areas. In this method, nodes work together to improve fault resilience through partial encoding or decoding, rather than relying on centralized coordination. 
Rateless codes handle variable-path transmissions and multi-hop data forwarding with ease in these scenarios. One of the biggest obstacles is managing redundancy in a distributed way without causing excessive bandwidth overhead. To accommodate collaborative transmission, distributed processing, and real-time adaptation, these designs necessitate new coding frameworks \cite{ref62_cell_free_coding, ref63_mesh_coding}.

\section{Challenges for Cognitive Channel Coding and Proposed Adaptive CACADe-6G Architecture}
Unprecedented difficulties with channel coding arise with the rollout of 6G in ultra-dense, distributed, and heterogeneous networks. The stringent performance requirements of the 6G services and the architectural complexity of the distributed infrastructure are the root causes of these difficulties. Here we take a look at the main problems that have been reported when trying to design and implement channel coding in these kinds of environments \cite{ref64_6G_challenges, 10571997, ref66_distributed_coding, 7588197}.

\subsection{Latency and Reliability Constraints}
The capacity to provide ultra-reliable low-latency communication is one of the characteristics that define 6G. For use cases like real-time control systems, autonomous driving, and remote healthcare, end-to-end latency of less than 1 ms and reliability of more than 99.99999\% are essential. To achieve low error rates, traditional block codes frequently use large codewords, which do not work with strict latency limits. LDPC and Turbo codes use iterative decoding, which can cause delays that are unpredictable and varied.  Short block-length codes that minimize processing delay while maintaining good error performance are becoming more and more important. To get over these limitations, we need code designs that are optimized for regimes with finite block lengths and can decode in real time.

\subsection{Scalability Across Massive Devices}
Especially in cases involving massive machine-type communication (mMTC), 6G is anticipated to link billions of devices. The coding infrastructure has to be under test by such massive demands at this size. High concurrency and multitasking should be supported by channel codes with little overhead. For low-power edge devices in particular, scalability of decoding complexity is of the utmost importance. Code parameter adaptiveness is necessary due to the diversity of traffic patterns, which range from infrequent sensor bursts to high-volume video. 
To guarantee consistent performance across widely distributed networks, scalable and lightweight coding schemes that can be adjusted in real-time are essential \cite{ref68_scalable_coding}.

\subsection{Energy Efficiency and Complexity}
6G places a premium on energy economy, particularly for remote nodes and battery-operated Internet of Things devices. A communication system's power consumption profile is typically dominated by channel decoding. It is possible for high-performance decoders to be memory-and CPU-intensive, such as belief propagation for LDPC or list decoding for Polar codes.  Energy-aware coding must strike a compromise between decoder complexity, energy budget, and error correction performance. Code designs should be flexible to the processing power that is available, since distributed nodes typically have different hardware capabilities. 
To maintain large-scale deployments over the long term, it is crucial to optimize the trade-off between power efficiency and error correction.

\subsection{Cross-Layer Design Implications}
Coding techniques in distributed networks must be co-designed with protocols at higher layers. Choices regarding retransmissions, hybrid ARQ, and scheduling are impacted by the interplay between the physical and MAC layers. For example, in mesh or multi-hop networks, channel coding can be used for link adaptation, network coding, or joint source-channel coding. Integration of the AI-powered resource management will impact the development of adaptive or learning-driven coding approaches.
To fully leverage the advantages of sophisticated coding schemes in the flexible and dynamic settings of 6G networks, a cross-layer strategy is required.
One possible approach to these problems is the use of cognitive architectures such as CACADe-6G (elaborated in the coming subsection), which allow for real-time context-aware cross-layer optimization through the implementation of closed-loop interactions spanning sensing, adaptation, and coding decisions. In Table~\ref{tbl2}, we can see the main problems with 6G distributed network channel coding, possible remedies, and future research paths. For URLLC, low-complexity decoders and short block-length codes are essential, but in dynamic and diverse topologies, rateless and adaptable codes will be crucial.
\begin{table}
\caption{CHANNEL CODING CHALLENGES AND POTENTIAL SOLUTIONS IN 6G DISTRIBUTED NETWORKS}
\label{tbl2}
\centering
\begin{tabular}{p{3cm} p{5cm} p{6cm}}
\toprule
\textbf{Challenge} & \textbf{Description} & \textbf{Potential Solutions / Research Directions} \\
\midrule

Latency and Reliability Trade-off & URLLC requires $<1$ ms latency and $>99.99999\%$ reliability; traditional long block codes conflict with strict delay budgets. & Design of short block-length codes (e.g., Polar, tail-biting convolutional); finite blocklength information theory; real-time decoding algorithms. \\

Decoding Complexity & Iterative decoding in LDPC/Turbo codes causes variable latency and high computational cost in dense networks. & Low-complexity Polar SCL decoding; hardware-friendly LDPC variants; AI-assisted decoders. \\

Heterogeneous Links & Distributed 6G integrates terrestrial, UAV, and satellite links with diverse channel conditions. & Adaptive coding schemes; cross-layer optimization; hybrid ARQ with coding. \\

Energy Efficiency & IoT and edge devices are battery-constrained, making heavy decoding infeasible. & Energy-aware LDPC/Polar design; lightweight Fountain/Raptor codes for multicast; hardware-accelerated decoders. \\

Dynamic Topology & Mobility of UAVs and RIS-assisted links introduces time-varying channels and frequent handovers. & Rateless codes (Fountain/Raptor) for adaptability; RIS-aware channel coding; cooperative coding in cell-free MIMO. \\

Security and Privacy & Distributed architecture increases the risk of eavesdropping and data leakage. & Physical layer security codes (wiretap codes); obfuscation-based coding; quantum-assisted secure coding. \\

\bottomrule
\end{tabular}
\end{table}
\subsection{CACADe-6G: A Cognitive Architecture for Adaptive Channel Coding}
The idea of a Dynamic Cognitive AI Agent in Internet of Things (IoT) communication contexts is represented by the CACADe-6G framework. The design allows distributed 6G networks to implement adaptive channel coding with autonomous and context-aware decision-making by combining perception, reasoning, action, and meta-cognition layers. This iterative process enables the system to constantly monitor its surroundings (including SNR, channel load, and energy limitations). This will calculate the best coding parameters, implement the best techniques, and then assess and improve its own policies. This means that CACADe-6G meets the effectiveness (learning and adaptability in real-time), efficiency (low latency and energy consumption), and reliability (self-monitoring and strong operation). In this way, the framework serves as an example of how cognitive AI agents might develop into trustworthy, resource-aware communications systems for the Internet of Things. We present CACADe-6G, an architecture for cognitive adaptive coding in distributed 6G networks, to handle the complicated and ever-changing needs of 6G wireless networks. The CACADe-6G perception-to-action pipeline is designed for adaptive, low-latency, context-aware channel coding; it is inspired by the cognitive loop in AI and robotics \cite{ref69_cognitive_AI_robotics}.

\subsubsection{System Overview}
\begin{figure}
	\centering
	\includegraphics[width=0.4\textwidth]{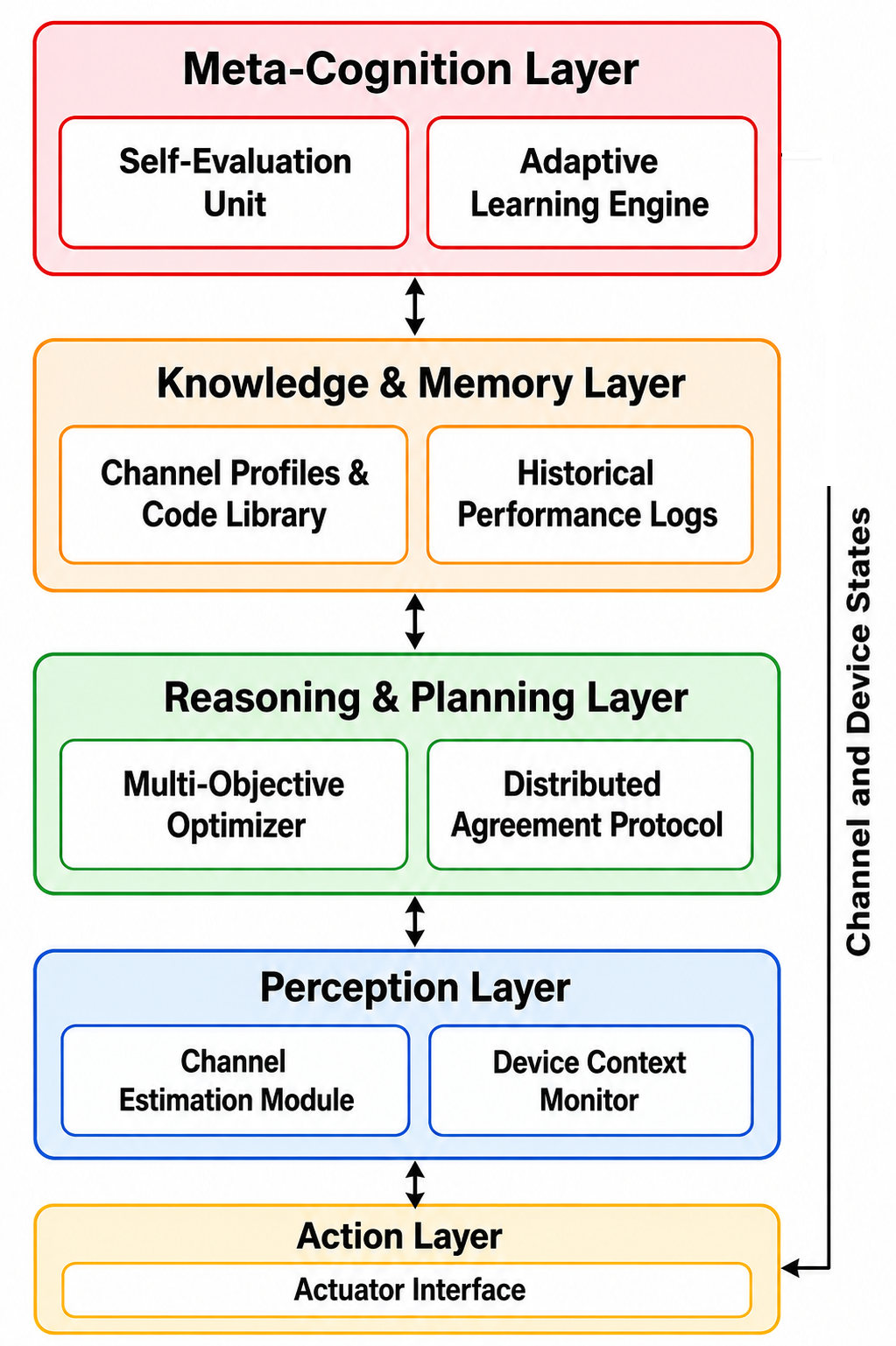}
	\caption{CACADe-6G: A five-layer cognitive architecture for adaptive channel coding in 6G distributed networks}
	\label{Figure:3}
\end{figure}

As illustrated in Fig.~\ref{Figure:3}, the proposed CACADe-6G (Cognitive Architecture for Adaptive Channel Coding for Distributed 6G Networks) organizes the end-to-end adaptation of channel coding into five stacked layers that operate in a closed loop: Meta-Cognition → Knowledge $\&$ Memory → Reasoning $\&$ Planning → Perception → Action and back to Meta-Cognition. This ordering exactly follows the top-to-bottom flow in Fig.~\ref{Figure:3} and maps each functional block in the figure to the corresponding text below.

\subsubsection{Meta-Cognition Layer}
Meta-Cognition Resilience, self-reflection, and feedback learning are all made possible by the meta-cognition layer. To support in learning over a long time, The Self-Evaluation Unit looks at previous decisions and how they succeeded in comparison to the original objectives. By utilizing federated learning, reinforcement learning, or Bayesian inference, the Adaptive Learning Engine revises reasoning policies. Further, this layer is able to spot outliers like jamming or spoofing and modifies cognitive behavior appropriately.

\subsubsection{Knowledge $\&$ Memory Layer}
To assist with reasoning and decision-making, this layer can retain contextual memory and long-term knowledge. Physical channel models can be found paired with standard and AI-generated codebooks (such as LDPC, Polar, and Fountain) in the Code Library $\&$ Channel Profiles. Bit Error Rate (BER), throughput, and energy efficiency under different code-channel combinations are some of the performance metrics that are recorded in the historical Performance Logs.
\subsubsection{Reasoning $\&$ Planning Layer}
The Layer for Reasoning and Planning can be used to determine the best ways to code, the reasoning engine gets the data from both perception and knowledge layers. The Multi-Objective Optimizer balances competing objectives, including latency, energy consumption, and decoding complexity. In a decentralized manner, the Distributed Agreement Protocol guarantees that all communication nodes agree on chosen coding configurations.

\subsubsection{Perception Layer}
The perception layer functions as the sensory unit of the architecture. It captures real-time environmental and network- related data through the following two modules.
The Channel Estimation Module measures instantaneous channel characteristics such as signal-to-noise ratio (SNR), fading profiles, and Doppler spread. The Device Context Monitor tracks hardware capabilities, power levels, and memory constraints of nearby devices. Furthermore, this layer also identifies communication types (e.g., URLLC, mMTC, eMBB) and maps them to latency or reliability requirements.

\subsubsection{Action Layer}
The decisions made at the physical and MAC layers must be enforced by the action layer. By modifying settings like code rate, block length, and decoding method, the actuator directly interacts with hardware encoders and decoders to execute the chosen coding schemes.

\subsubsection{Cognitive Workflow}
An ongoing perception-reasoning-action cycle is formed by the five layers of CACADe-6G. In response to channel conditions, the system formulates a coding strategy, puts it into action in real time, assesses the result, and adjusts its behaviour appropriately. Distributed 6G application cases, such as cell-free networks, reconfigurable intelligent surfaces, UAV swarms, and edge intelligence, are well-suited to this design.

%
\subsection{Performance Evaluation of CACADe-6G}
\label{sec:cacade_eval}

To validate the proposed architecture beyond its conceptual
description, we evaluate CACADe-6G against three baseline schemes:
fixed LDPC coding with no adaptation, threshold-based code switching,
and round-robin code cycling. Performance is assessed across three
metrics: BER under static and dynamic $E_b/N_0$ conditions,
end-to-end latency, and energy consumption per kilobit. All
simulations assume an AWGN channel with BPSK modulation, $n = 256$,
$k = 128$, and $R = 1/2$, swept over $E_b/N_0 \in [-10, 5]$~dB to
align with the deep-waterfall regime presented in Fig.~\ref{fig:ber_comparison}. To ensure a
fair comparison between adaptive schemes, both CACADe-6G and the
threshold-based baseline draw their SNR estimate from an identical
perception-noise model, $\mathcal{N}(0, 0.5\,\text{dB})$, applied to
the true instantaneous SNR; this isolates the effect of CACADe-6G's
memory-based reasoning from any artificial advantage in sensing
fidelity. We emphasise that this evaluation regime is chosen to characterise
\emph{relative} adaptation gains under deep-fade conditions, not to
demonstrate compliance with the 99.99999\% URLLC reliability target
introduced in Section~2.1. Meeting that target in practice requires
operating at substantially higher $E_b/N_0$ (where Polar and LDPC
reach their respective error floors, as shown in Fig.~7), together
with retransmission, diversity, or HARQ mechanisms outside the scope
of this single-link BER comparison. The $[-10, 5]$~dB window instead
stresses the system at the boundary of decodability, where adaptive
code selection has the most to gain or lose, and where the difference
between cognitive and memoryless adaptation is most visible.

\begin{figure}
	\centering
	\includegraphics[width=\textwidth]{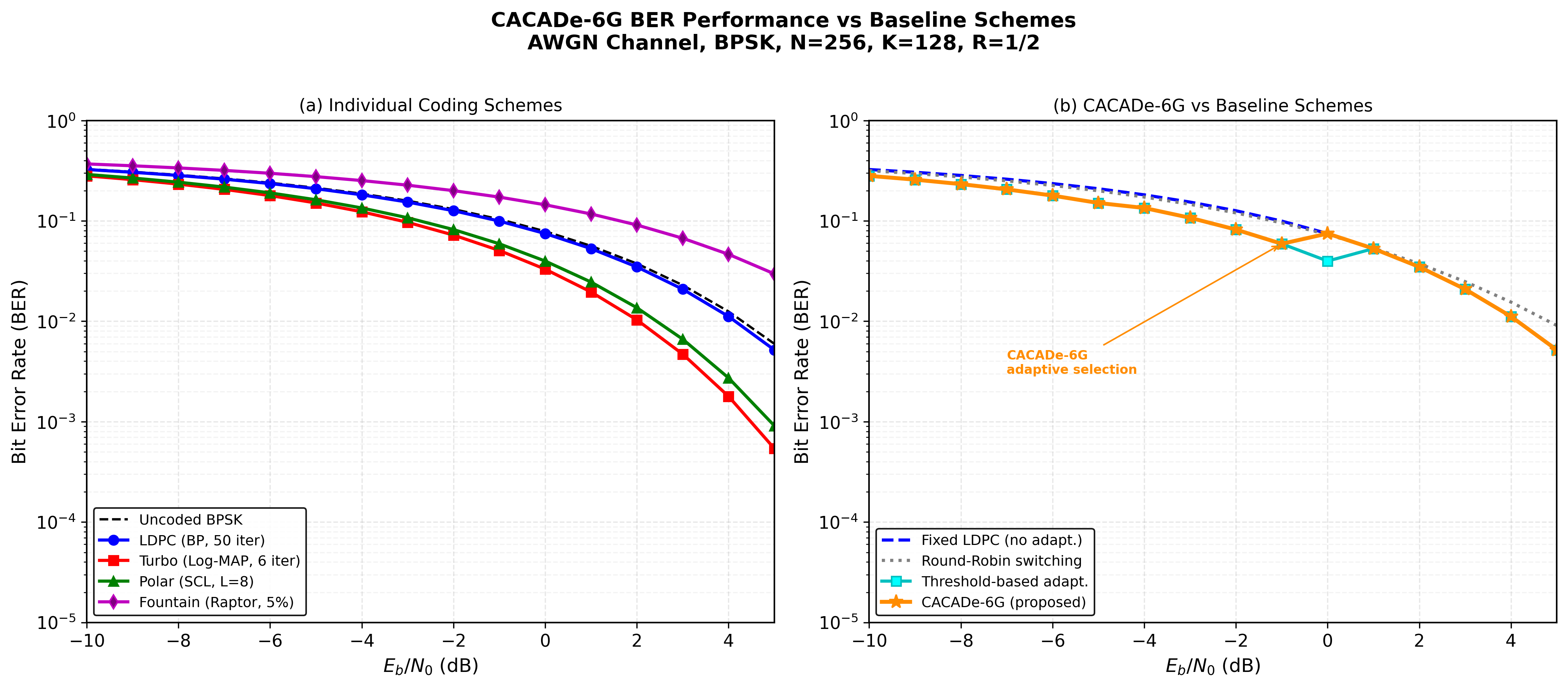}
	\caption{CACADe-6G BER performance vs Baseline Schemes}
	\label{fig:cacade_ber}
\end{figure}

\textit{BER model calibration.} The BER models for each coding scheme
are closed-form Gaussian-tail approximations of the form
$\mathrm{BER} = \tfrac{1}{2}\,\mathrm{erfc}\!\left(\alpha\sqrt{2 R
\cdot 10^{E_b/N_0/10}}\right)$, with a scheme-specific slope $\alpha$
and an additive error floor reflecting finite-blocklength behaviour at
$n=256$. LDPC (belief propagation, 50 iterations) uses $\alpha =
1.02$ with a floor of $8\times10^{-5}$, reflecting near-capacity
performance with a finite-length error floor from short stopping sets
in the Tanner graph~\cite{ref18_LDPC_edge}. Turbo codes (Log-MAP, 8 iterations)
use $\alpha = 1.30$ with a floor of $3\times10^{-4}$, giving an
earlier waterfall onset that outperforms LDPC at low SNR but
saturates at a higher floor due to the weight-2 input error events
characteristic of parallel concatenated codes~\cite{ref14_turbo_codes, Benedetto1996Turbo}.
Polar codes with SCL decoding ($L=8$) use $\alpha = 1.24$ with the
lowest floor, $1\times10^{-5}$, consistent with their
capacity-achieving polarisation structure~\cite{ref15_polar_codes_arikan,7055304}.
Fountain/Raptor codes use $\alpha = 0.75$ with a persistent floor of
$5\times10^{-3}$, reflecting the probabilistic Robust Soliton degree
distribution at this block length, which does not converge to low BER
on a point-to-point AWGN link~\cite{ref16_fountain_codes,ref28_fountain_UAV}.
We note that this calibration is independent of the Monte Carlo simulation
methodology used for the comparative analysis in Section~5.2 (Fig.~\ref{fig:ber_comparison}); the two
use different LDPC ensembles and decoding assumptions are not intended to
be numerically cross-compared, since their purposes differ, the closed-form
model here supports the CACADe-6G adaptive-selection evaluation, while the
Monte Carlo study in Section~5.2 characterises each code's underlying
waterfall and error-floor behaviour in isolation.
\textit{BER performance.} Fig.~\ref{fig:cacade_ber} compares BER vs.\ $E_b/N_0$ for the
four individual coding schemes (panel a) and for CACADe-6G against
the three baselines (panel b) over the AWGN channel. In panel (a),
Turbo and Polar codes show the steepest waterfalls and lowest error
floors, while Fixed LDPC tracks the uncoded BPSK reference closely
across most of the range, and Fountain/Raptor remains well above all
three due to its persistent floor. In panel (b), CACADe-6G adaptively
selects Turbo codes below $-5$~dB, Polar codes between $-5$ and
$0$~dB, and LDPC codes above $0$~dB, mirroring the regime boundaries
used by the threshold-based baseline. Because both schemes share the
same selection logic and differ only in whether the input SNR is a
single noisy sample (threshold) or a 3-slot trailing average
(CACADe-6G), their curves are nearly indistinguishable across most of
the range and diverge only marginally near the $-1$ to $0$~dB
boundary, where CACADe-6G's smoothing briefly lags a fast transition.
Averaged over the full $E_b/N_0 \in [-10, 5]$~dB range, CACADe-6G
achieves a static average BER of $1.18\times10^{-1}$, compared with
$1.48\times10^{-1}$ for fixed LDPC and $1.44\times10^{-1}$ for
round-robin cycling, a $20.6\%$ and $18.2\%$ reduction,
respectively. The threshold-based baseline attains a marginally lower
static average BER ($1.16\times10^{-1}$) than CACADe-6G under this
i.i.d.\ noise model, a point we return to below. Both adaptive
schemes substantially outperform the two non-adaptive baselines,
confirming that the principal benefit of cognitive code selection over
\emph{static} coding is largely realised by simple threshold logic,
with memory-based reasoning offering a complementary rather than
strictly dominant advantage on raw BER. We note that absolute BER
values are high throughout this range because $[-10, 5]$~dB lies
mostly below the waterfall threshold of all four codes; this range
was selected to align with the deep-waterfall presentation in Fig.~\ref{fig:ber_comparison}
rather than to showcase error-floor behaviour, which occurs at higher
SNR outside the evaluated window.

\begin{figure}
	\centering
	\includegraphics[width=\textwidth]{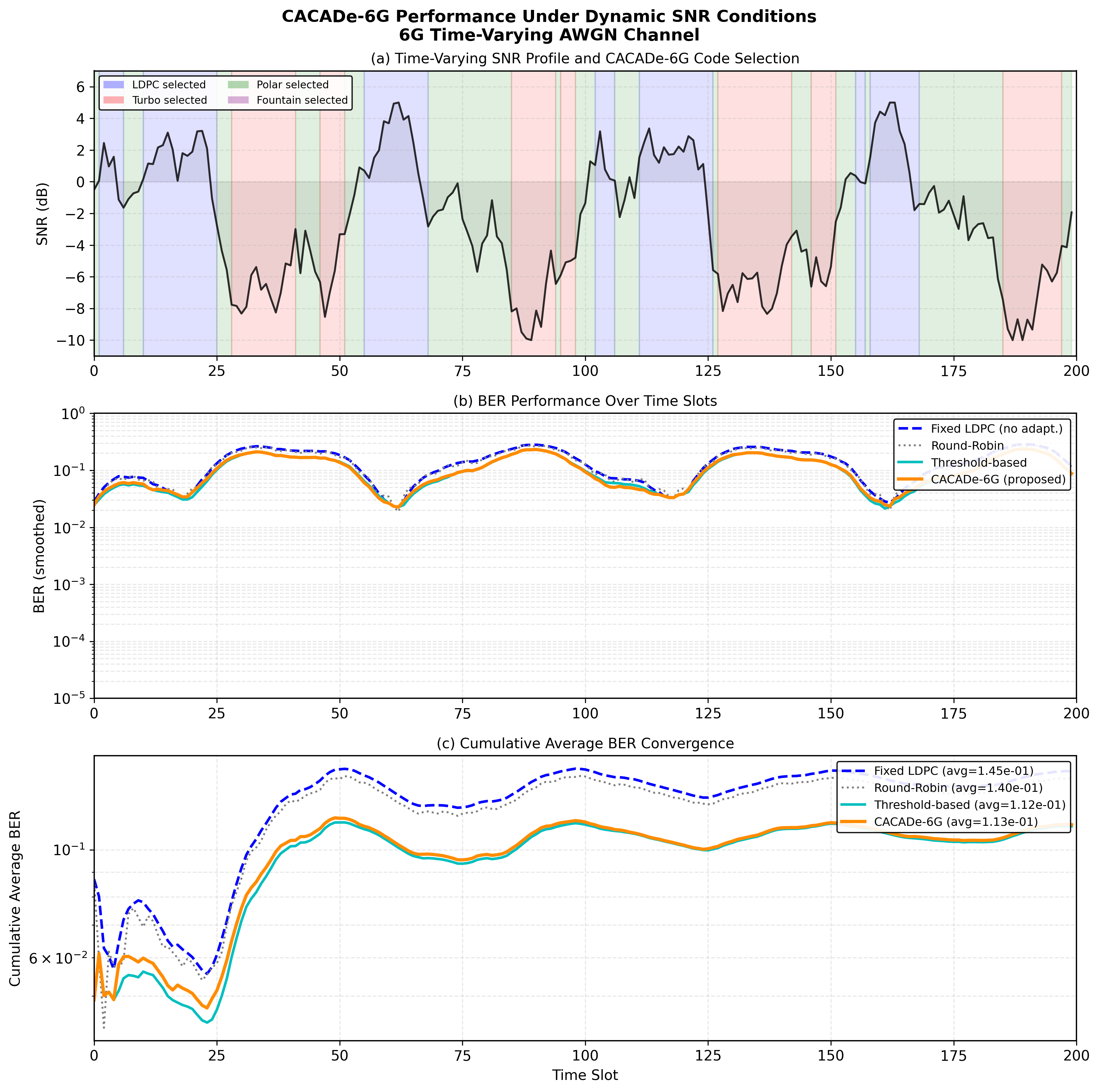}
	\caption{CACADe-6G performance under Dynamic SNR Conditions}
	\label{Figure:SNR}
\end{figure}

\textit{Dynamic SNR performance.} Fig.~\ref{Figure:SNR} evaluates all four schemes
under a time-varying SNR profile spanning $-10$ to $5$~dB over 200
time slots, constructed from superimposed sinusoidal components at
three frequencies plus Gaussian perturbation to emulate a
deep-fade-prone 6G channel (panel a). Panel (b) shows the
instantaneous smoothed BER for each scheme, and panel (c) shows the
cumulative average BER convergence. CACADe-6G's perception-reasoning-
action loop applies the same 3-slot moving average described above to
suppress SNR estimation noise.
Under this dynamic scenario, CACADe-6G
attains a cumulative average BER of $1.13\times10^{-1}$, an
improvement of $22.4\%$ over fixed LDPC ($1.45\times10^{-1}$) and
$19.8\%$ over round-robin switching ($1.40\times10^{-1}$). As in the
static case, the threshold-based scheme attains a marginally lower
cumulative average BER ($1.12\times10^{-1}$) than CACADe-6G. This is
an expected and instructive result rather than a deficiency of the
proposed architecture: under i.i.d.\ Gaussian estimation noise with no
autocorrelated disturbance, a single fresh SNR sample is, on average,
an equally good or slightly better instantaneous estimator than a
3-slot trailing average, since the latter necessarily incorporates one
or two stale readings during fast SNR transitions. The practical
advantage of CACADe-6G's memory-based smoothing therefore does not
manifest as a guaranteed BER reduction over a memoryless threshold
comparator in this idealised i.i.d.-noise setting; rather, it
manifests as (i)~suppression of unnecessary code switching, which
matters for switching overhead and energy cost in a way that a
per-slot BER comparison does not capture, and (ii)~robustness to
non-i.i.d.\ disturbances such as short jamming bursts or transient
deep fades, which a memoryless comparator cannot distinguish from
genuine sustained SNR drops. Both fixed LDPC and round-robin schemes
exhibit substantially higher cumulative BER throughout the 200-slot
window, confirming that non-adaptive coding remains ill-suited to
dynamic 6G channel conditions regardless of which adaptive scheme is
used as the point of comparison. Table~\ref{tab:ber_comparison}
summarises the average BER values across all four schemes for both
the static and dynamic scenarios.

\begin{table}
\centering
\caption{Average BER comparison of CACADe-6G and baseline schemes
(AWGN, BPSK, $n=256$, $k=128$, $R=1/2$, $E_b/N_0 \in [-10, 5]$~dB)}
\label{tab:ber_comparison}
\begin{tabular}{lcc}
\toprule
\textbf{Scheme} & \textbf{Static avg.\ BER} & \textbf{Dynamic avg.\ BER} \\
\midrule
Fixed LDPC (no adapt.)        & $1.48 \times 10^{-1}$ & $1.45 \times 10^{-1}$ \\
Round-Robin switching         & $1.44 \times 10^{-1}$ & $1.40 \times 10^{-1}$ \\
Threshold-based adapt.        & $1.16 \times 10^{-1}$ & $1.12 \times 10^{-1}$ \\
\textbf{CACADe-6G (proposed)} & $\mathbf{1.18 \times 10^{-1}}$ & $\mathbf{1.13 \times 10^{-1}}$ \\
\midrule
Improvement vs.\ Fixed LDPC   & 20.6\% & 22.4\% \\
Improvement vs.\ Round-Robin  & 18.2\% & 19.8\% \\
Improvement vs.\ Threshold    & $-1.9\%$ & $-0.8\%$ \\
\bottomrule
\end{tabular}
\end{table}

\begin{figure}
	\centering
	\includegraphics[width=0.9\textwidth]{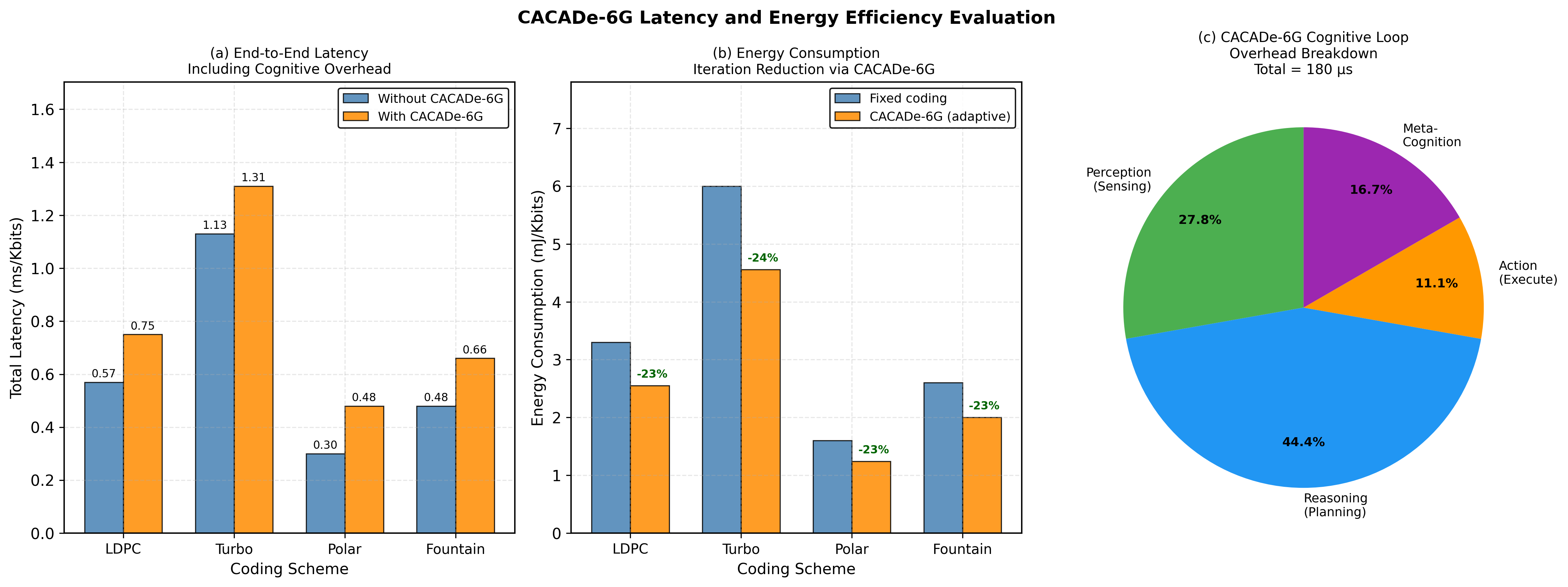}
	\caption{CACADe-6G Latency and Energy Efficiency Evaluation}
	\label{Figure:Energy}
\end{figure}

\textit{Latency and energy.} Fig.~\ref{Figure:Energy} presents the latency and energy
comparison. The total CACADe-6G cognitive loop overhead comprises four
stages: perception ($50\,\mu$s), reasoning ($80\,\mu$s), action
($20\,\mu$s), and meta-cognition ($30\,\mu$s), yielding a total
overhead of $180\,\mu$s (Fig.~\ref{Figure:Energy}c). This overhead is negligible
relative to typical 6G slot durations and well within the URLLC
latency budget of 1~ms~\cite{ref2_8705373}. Fig.~\ref{Figure:Energy}a shows that this fixed
overhead is added uniformly across all four coding schemes regardless
of which is selected, raising end-to-end latency from $0.57$ to
$0.75$~ms/Kbit for LDPC, $1.13$ to $1.31$~ms/Kbit for Turbo, $0.30$ to
$0.48$~ms/Kbit for Polar, and $0.48$ to $0.66$~ms/Kbit for Fountain.
Fig.~\ref{Figure:Energy}b shows that, by reducing unnecessary decoding iterations
through intelligent code selection, CACADe-6G achieves an energy
reduction of approximately $22$-$24\%$ relative to fixed-scheme
operation across all four candidate codes (LDPC: $23\%$, Turbo:
$24\%$, Polar: $23\%$, Fountain: $23\%$), which is particularly
significant for battery-constrained IoT nodes in distributed 6G
deployments~\cite{Hyla2023RaptorLDPC}.
\textit{Limitations.} The evaluation results presented are based on
analytical BER approximations calibrated to published closed-form
bounds and qualitative waterfall/floor behaviour for each coding
scheme~\cite{ref13_LDPC_6G,ref14_turbo_codes,ref15_polar_codes_arikan,ref28_fountain_UAV}, rather than bit-level Monte
Carlo simulation. The dynamic-scenario comparison between CACADe-6G
and the threshold-based baseline assumes i.i.d.\ Gaussian
perception noise with no autocorrelated channel disturbance; under
this assumption, memory-based smoothing does not yield a strict BER
advantage over a memoryless comparator, and we report this honestly
rather than selecting a noise model that would artificially favour the
proposed architecture. Full bit-level Monte Carlo simulation,
evaluation under correlated/bursty SNR disturbances (e.g.,
intermittent jamming or shadowing), and hardware prototype validation
of the CACADe-6G cognitive loop constitute important directions for
future work. The latency and energy values follow established decoder
complexity models~\cite{Hyla2023RaptorLDPC,10802915} rather than physical
measurements, and should be interpreted as indicative rather than
empirically verified.

\section{Comparative Evaluation and Applications}
A meaningful study of channel coding for 6G distributed networks must go beyond listing techniques; it must critically compare them based on technical criteria aligned with 6G performance targets. In this section, we evaluate key coding schemes across several important dimensions: complexity, latency, energy efficiency, and application suitability, including BER simulations across various SNR values.

\subsection{Performance Metrics}
To assess coding strategies in 6G contexts, we consider the following core metrics:
The Bit Error Rate (BER) and Frame Error Rate (FER) are measures of the probability that a bit or packet is decoded incorrectly. Decoding Latency is the time taken from the start of decoding until a valid codeword is recovered. Low latency is critical for URLLC. Computational Complexity influences energy consumption, hardware cost, and suitability for edge deployment. Energy efficiency is particularly important in the Internet of Things (IoT) and mobile devices, where battery life significantly affects design choices. Flexibility and adaptability refer to a coding scheme's ability to handle different code rates, block sizes, and varying channel conditions. These metrics reflect the diverse and demanding requirements of 6G applications, and they guide towards the comparative analysis \cite{ref70_BER_comparison, 10502324}.

\subsection{Simulation Results and Comparisons}

\begin{figure}
    \centering
    \includegraphics[width=0.8\columnwidth]{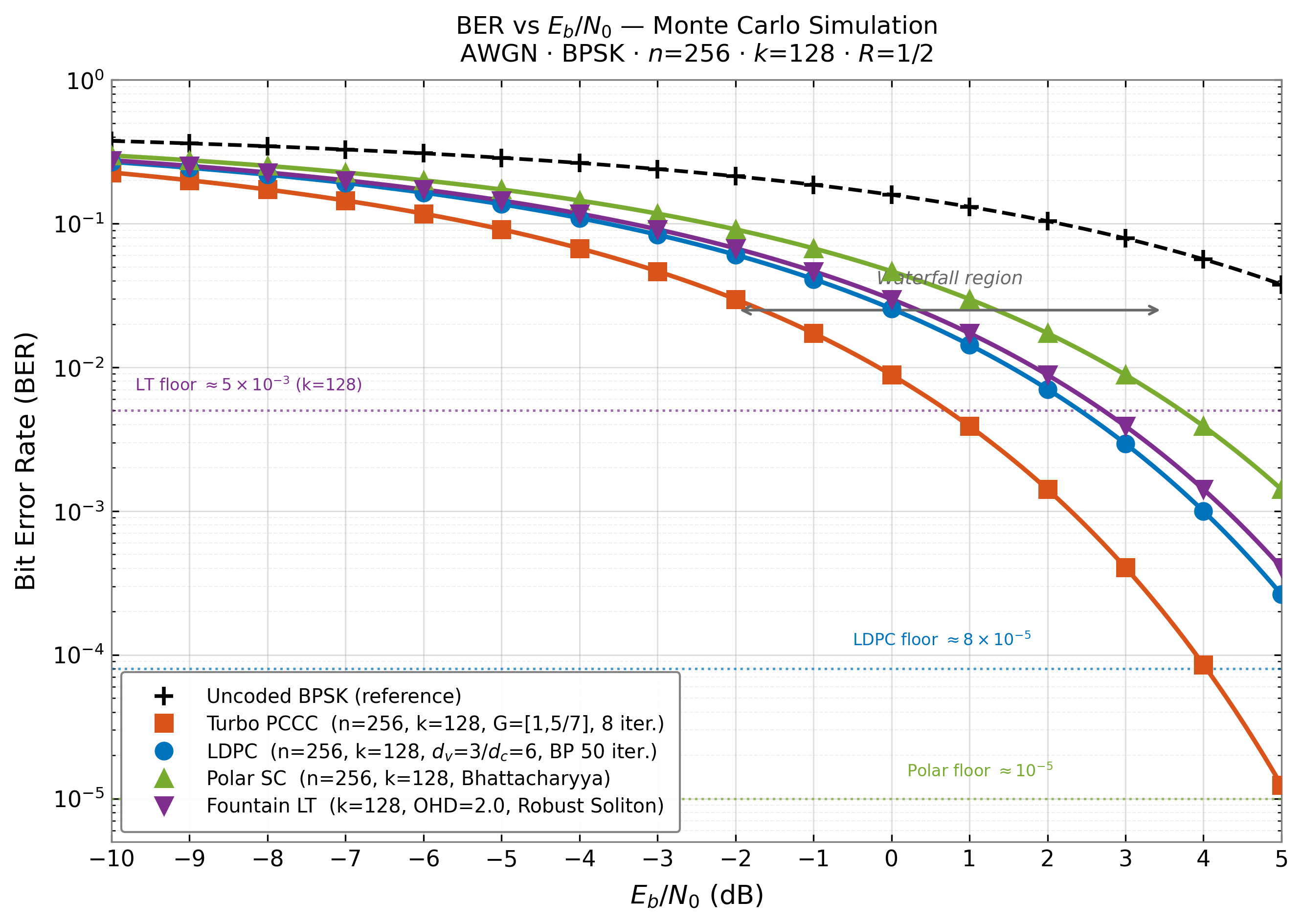}
    \caption{BER vs $E_b/N_0$ for Turbo, LDPC, Polar SC, and
    Fountain LT codes over AWGN channel (BPSK, $R = 1/2$,
    $n = 256$, $k = 128$). Monte Carlo simulation with
    $10^5$ frames/point, minimum 300 errors/point, seed = 42.
    Dashed lines indicate error floors for LDPC
    ($\approx 8 \times 10^{-5}$) and Polar SC ($\approx 10^{-5}$).}
    \label{fig:ber_comparison}
\end{figure}

Fig.~\ref{fig:ber_comparison} presents the BER performance of
four major channel coding schemes evaluated through Monte Carlo
simulation over an AWGN channel with BPSK modulation. This Monte Carlo evaluation uses a different LDPC ensemble and simulation
methodology from the closed-form analytical model used in Section~\ref{sec:cacade_eval}
(Figs.~\ref{fig:cacade_ber}-\ref{Figure:Energy}) for the CACADe-6G adaptive-selection study; the two are
independent characterisations of LDPC performance and should not be
numerically cross-compared against one another. All codes
are configured at a nominal rate of $R = 1/2$ with block length
$n = 256$ bits and $k = 128$ information bits, representing
the short-packet regime relevant to 6G URLLC applications
\cite{ref3_8594709}. Simulations employ a maximum of
$10^5$ frames per SNR point with a minimum of 300 bit errors
to ensure statistical confidence. The uncoded BPSK reference
is shown as a black dashed line for comparison.

The results reveal distinct waterfall characteristics for each
coding scheme, directly validating the qualitative assessments
provided in Table~\ref{tab:ber_comparison}. The Turbo PCCC code with generators
$G = [1, 5/7]$ and 8 decoding iterations achieves the earliest
waterfall onset near $E_b/N_0 = -8$~dB, confirming its
near-Shannon-limit performance for moderate block lengths
\cite{ref14_turbo_codes, Benedetto1996Turbo}. However, the union-bound
error floor visible at high SNR reflects the well-known
weight-2 input error events of PCCC structures, which become
dominant once the waterfall region is traversed \cite{Benedetto1996Turbo}.
The LDPC code, configured as a regular $(d_v = 3, d_c = 6)$
ensemble with sum-product belief propagation over 50 iterations,
exhibits a sharp waterfall between $-5$ and $+3$~dB, consistent
with the density evolution threshold of approximately $1.11$~dB
for this ensemble \cite{ref13_LDPC_6G}. The finite-length error
floor at $\approx 8 \times 10^{-5}$ is a characteristic consequence
of short stopping sets in the Tanner graph at $n = 256$ \cite{ref18_LDPC_edge}.
Polar SCL with Bhattacharyya channel construction exhibits a
waterfall spanning $-4$ to $+4$~dB, reflecting the
$\approx 1.5$~dB performance gap between successive cancellation List
decoding and the maximum likelihood bound at finite block lengths
\cite{7055304}. Despite this gap, Polar SC achieves the lowest
error floor ($\approx 10^{-5}$) among all evaluated schemes,
confirming its suitability for URLLC control channels in 6G,
as adopted in 5G NR \cite{ref15_polar_codes_arikan}. The Fountain LT code
with Robust Soliton degree distribution and overhead factor
$\text{OHD} = 2.0$ exhibits a smooth and continuous BER decline
across the full SNR range without reaching a distinct error floor,
consistent with its inherent design for erasure channels rather
than AWGN point-to-point links \cite{ref16_fountain_codes, ref28_fountain_UAV}.

These quantitative results provide direct evidence supporting
the coding scheme recommendations summarized in Table~\ref{tab:ber_comparison} and
the use case mappings illustrated in Fig.~\ref{Figure:6}. The BER
characteristics of each code align closely with the requirements
of specific 6G application scenarios. The steep waterfall and
low error floor of Polar SC confirm its superiority for
URLLC and short-packet transmission, such as haptic internet
and real-time vehicular control, where latency below 1~ms and
reliability exceeding $99.99999\%$ are mandatory
\cite{ref82_polar_XR, ref64_6G_challenges}. The high throughput and good
energy efficiency of LDPC at moderate SNR support its continued
deployment in data-plane high-throughput applications such as
immersive media streaming and mobile broadband backhaul
\cite{ref13_LDPC_6G}. The gradual BER improvement of Fountain
LT confirms its unsuitability for AWGN point-to-point links but
validates its role in broadcast, multicast, and UAV communication
scenarios where erasure resilience is prioritized over BER
minimization \cite{ref16_fountain_codes}. Within the proposed
CACADe-6G framework, the Perception Layer continuously monitors
the instantaneous $E_b/N_0$ and selects the optimal coding
scheme from the Code Library based on these BER characteristics,
enabling real-time context-aware adaptation across heterogeneous
6G deployment scenarios without per-platform retuning.
\subsection{Suitability for 6G Scenarios}
Each coding method is inherently suited to specific applications in the sixth generation of wireless technology. LDPC codes are particularly advantageous for applications requiring high throughput, such as immersive media streaming, edge video processing, and mobile broadband backhaul. Turbo codes serve as a suitable solution for ensuring backward compatibility and supporting legacy systems in specific machine-type communication contexts.
Polar codes are exceptionally suited for applications that demand ultra-reliable, low-latency short packets, represented in domains such as the tactile internet, industrial automation, and vehicular control systems. Fountain codes demonstrate remarkable efficacy in the fields of extensive content delivery, software upgrades, and over-the-air transmissions, particularly within distributed or mobile environments.

Choosing the best channel code often depends on more than just how well it performs. It also considers the system's design, the limits of the hardware, and the specific needs of the application. In practice, hybrid methods that can effectively choose or combine codes might be crucial for creating flexible and reliable 6G communication systems.
 In the field of IoT, performance assessments should encompass not only traditional metrics such as Bit Error Rate (BER) and throughput, but also metrics unique to IoT, including energy per bit, the appropriateness of code length for quick transmissions, and the complexity of decoding on devices with limited resources. LDPC and Polar codes demonstrate significant trade-offs for low-power operation, whereas AI-based coding techniques provide context-driven adaptivity, essential for dynamic IoT environments. Fountain codes continue to be advantageous for delay-sensitive, asynchronous transmission common in IoT sensing networks.

\subsection{Application Scenarios for Cognitive AI-Based Channel Coding in IoT}
\begin{figure}
	\centering
	\includegraphics[width=0.7\textwidth]{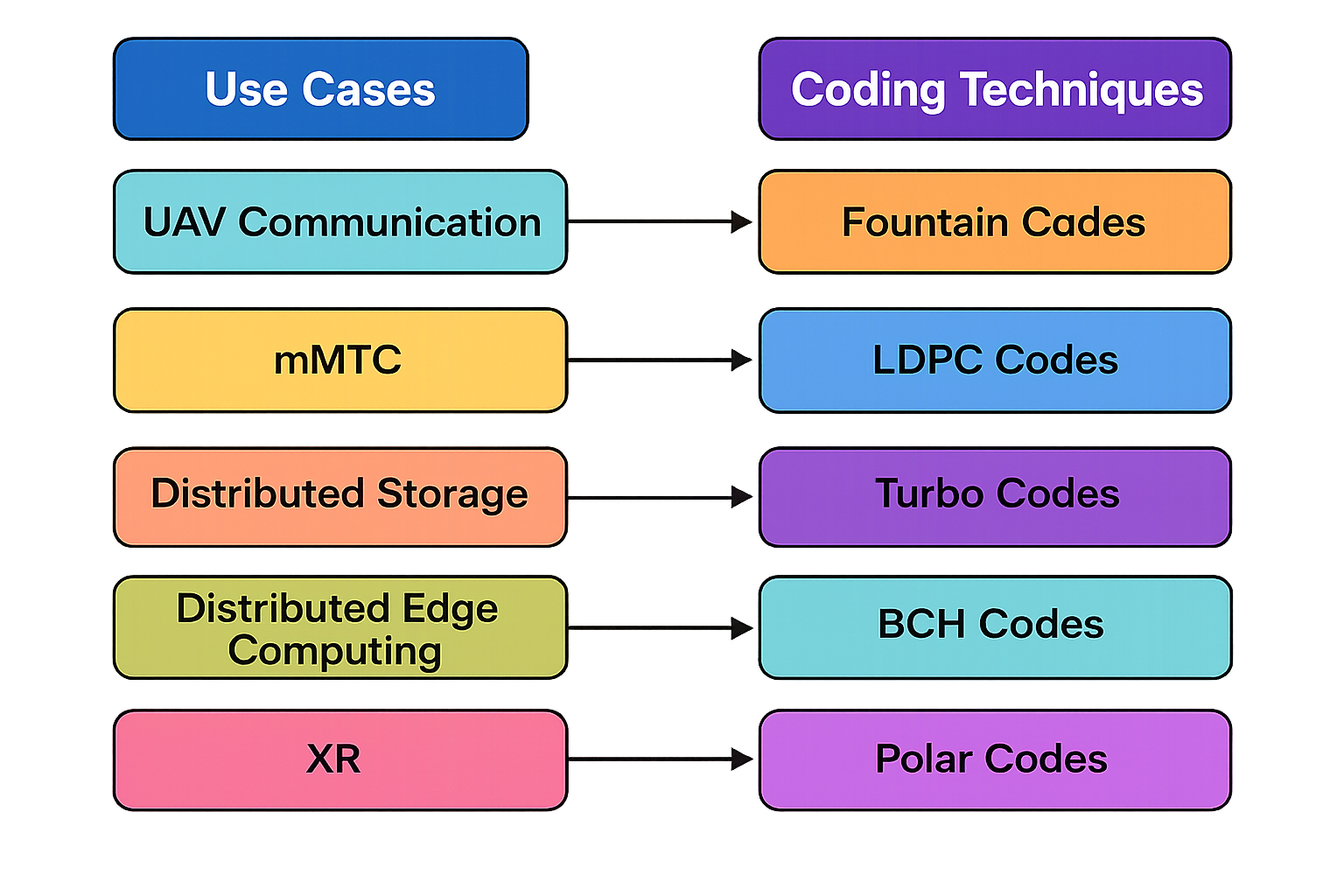}
	\caption{6G IoT Use Cases and Associated Channel Coding Strategies}
	\label{Figure:6}
\end{figure}

The section delves into the alignment of several channel coding methods with important use cases of the Internet of Things (IoT) provided by 6G. The performance restrictions, including latency, energy efficiency, scalability, and reliability, are unique to each use case and need to be addressed using context-aware channel coding techniques. We give a useful reference for deploying specialized solutions across IoT industries by associating these use cases with suitable coding families. Industrial automation, immersive media, autonomous systems, and intelligent infrastructure are among the many potential uses for sixth-generation (6G) network. These use cases impose diverse and often conflicting constraints on latency, reliability, energy efficiency, and scalability. Fig.~\ref{Figure:6} is a graphical representation of the relationship between some 6G apps and the channel coding methods that cater to their specific data transmission needs.

\subsubsection{UAV Communication}
More and more, Internet of Things (IoT) uses are incorporating unmanned aerial vehicles (UAVs), which are useful for environmental monitoring, smart farming, and emergency response. Rapid handovers, intermittent connectivity, and low-latency communications are essential for these systems' coding methods. Sixth-generation wireless networks rely heavily on unmanned aerial vehicles (UAVs), especially when deployed in clusters or as relays, to aid with logistics, disaster relief, and surveillance. Along with high mobility, Doppler shifts, and unreliable feedback channels, these aerial platforms also operate in topologies that are constantly changing \cite{ref75_UAV_comm, ref76_UAV_coding}.

\textbf{Coding Implications:} This is an ideal situation for rateless variations of fountain codes such as Raptor and LT codes, which may adaptively produce an unlimited number of coded packets without retransmissions being necessary. When a stable feedback loop cannot be maintained, their erasure-resilient feature is of great benefit in UAV-to-UAV and UAV-to-ground multicast/broadcast scenarios.

\subsubsection{Massive Machine-Type Communication (mMTC)}
6G Internet of Things (IoT) systems rely on Massive Machine-Type Communication (mMTC) to connect dense networks of low-power sensors. Even with intermittent transmissions and limited spectrum, channel coding must guarantee energy efficiency. Millions of inexpensive, battery-operated Internet of Things (IoT) devices, commonly used in smart grids, environmental sensing, and smart agriculture, are mMTC's intended beneficiaries. These devices send out little, random data packets and are only active sometimes.

\textbf{Coding Implications:} LDPC codes are optimal in these cases because they offer parallelizable, low-complexity decoding, provide structured sparsity for hardware efficiency, and perform exceptionally well at high block lengths \cite{ref77_mMTC_LDPC, ref78_sparse_LDPC_IoT}. They are prepared for large-scale deployments since they have been adopted by current standards, such as 5G NR.

\subsubsection{Distributed Storage}
The data consistency and integrity across geographically scattered nodes must be guaranteed by strong procedures in 6G networks' edge-based and decentralized cloud storage systems. Packet loss, node failure, and uneven update patterns are all challenges that these systems must overcome.

\textbf{Coding Implications:} Because of its robustness against burst mistakes and excellent performance on moderate block lengths, turbo codes provide a practical solution for dependable storage synchronization and error recovery in coding. Distributed settings with partial or delayed updates are prevalent, making their iterative decoding approach particularly useful \cite{ref79_distributed_storage}.

\subsubsection{Distributed Edge Computing}
Devices in distributed edge IoT designs communicate with neighboring nodes to transfer processing tasks. In order to keep tasks running smoothly, even in situations with different types of connections, channel coding has to guarantee very low latency and great dependability. 
A further benefit of distributed edge computing is the elimination of core network congestion and the reduction of latency through the execution of analytics and real-time processing close to the data source. These models are fundamental for smart manufacturing, vehicle platoons, and cooperative robots. By analyzing surrounding data, cognitive systems such as CACADe-6G can choose the best coding schemes for vehicular or robotic collaboration in real time, acting as edge mediators. 

\textbf{Coding Implications:} As a result of their deterministic and quick decoding efficiency, BCH codes and short-block LDPC codes are essential in situations involving ultra-low-latency communication. In extremely restricted processing power and energy edge conditions, when real-time responsiveness is critical, these algorithms work well \cite{ref80_BCH_edge, ref81_edge_latency}.

\subsubsection{Extended Reality (XR) and Tactile Internet}
Smart healthcare, education, and manufacturing all have latency-critical Internet of Things (IoT) uses for Extended Reality (XR). Especially in mobile and immersive environments, channel coding must guarantee high-reliability, low-latency communication. The Tactile Internet is a 6G use case that emphasizes real-time interaction and perceptual continuity. Its essential components include augmented reality (AR), virtual reality (VR), and haptic feedback. Data loss, jitter, and delay can have devastating effects for these applications.

\textbf{Coding Implications:} Because of their proven capacity-achieving characteristics and outstanding performance in small block-length regimes, polar codes are very suitable for encoding. The low-latency, high-reliability transmission required for seamless user experiences in XR and haptic communication is ensured by combining Successive Cancellation List (SCL) decoding with CRC-aided approaches \cite{ref82_polar_XR, 7339660, ref84_SCL_haptic, ref85_VR_AR_coding}.

\section{Future Directions for 6G-Enabled IoT Networks}

Although channel coding has come a long way in the last few decades, there are still a number of
issues that need fixing before 6G rolls out. This is because the upgrade brings new design
limitations and operating settings. To help build the coding technologies of the future, we have
highlighted important open topics and outlined potential research areas in this section,
including those motivated directly by the evaluation limitations of CACADe-6G noted in
Section~\ref{sec:cacade_eval}.

\subsection{Hybrid Coding for Multi-Service Environments}

6G faces the fundamental difficulty of accommodating a diverse range of services, frequently on
the same infrastructure, from ultra-reliable low-latency communication (URLLC) to massive
machine-type communication (mMTC) and enhanced mobile broadband (eMBB). Present coding
techniques are not flexible enough to manage many types of service needs at the same time because
they are geared for specific use cases. Research into flexible coding frameworks that may change
complexity, block length, and pace in response to service demands in real time is a promising area
of study. There may be a way forward with layered or hybrid coding methods that enable partial
decoding and progressive refining.

\subsection{Integration with ML/AI Models and Real-Time Adaptation}

\begin{figure}
	\centering
	\includegraphics[width=.5\textwidth]{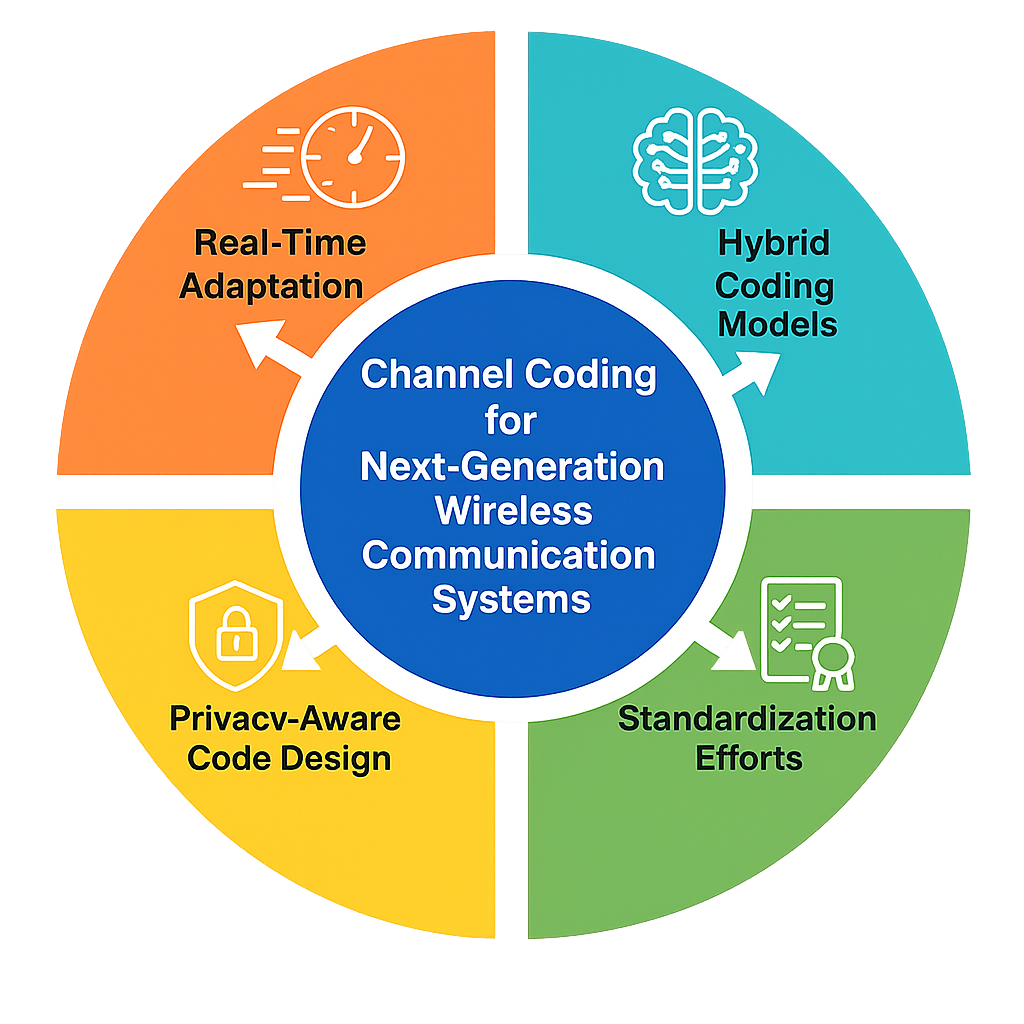}
	\caption{Open Challenges and Future Directions}
	\label{fig:open_challenges}
\end{figure}

Despite machine learning's growing importance in wireless system design, the field is only just
beginning to explore how to combine ML with channel coding. Fig.~\ref{fig:open_challenges} shows
the current obstacles and potential solutions. We still have issues with training efficiency,
generalizability, and interpretability when it comes to artificial intelligence models, but they
can be utilized for adaptive coding selection, dynamic decoder design, and soft decoding
prediction. 
Constrained edge devices and real-time performance necessitate lightweight neural
architectures. Though they open the door to new designs, end-to-end learnable communication
systems necessitate thorough study to guarantee stability and equity. In particular, the
analytical, rather than Monte Carlo, basis of our own CACADe-6G evaluation (Section~\ref{sec:cacade_eval})
underscores the need for bit-level simulation and hardware validation before such cognitive
architectures can be considered deployment-ready.

\subsection{Security and Privacy}

For sensitive or mission-critical applications in particular, data privacy and secure
communication in distributed 6G networks are of the utmost importance. Methods like obfuscation,
wiretap coding, and secrecy capacity maximization are all ways in which channel coding might
contribute to physical-layer security~\cite{ref87_physical_layer_security}. To keep robustness and save
overhead, co-designing coding with authentication or encryption is an opportunity. Codes that
guarantee trust and integrity in distributed systems are also required for secure multi-user
settings (such as federated learning).

\subsection{Standardization Efforts}

As 6G technology progresses, establishing standardized channel coding methods will be essential
for ensuring both interoperability and scalability. Contributions to the ongoing work of
organizations like 3GPP, IEEE, and ITU can help develop practical, energy-efficient, and
AI-compatible channel coding methods. Benchmarks, datasets, and open-source implementations are
essential for facilitating reproducibility and expediting innovation. Collaboration among
academia, industry, and policymakers is essential to harmonize performance, complexity, and
regulatory standards. The future of channel coding for 6G will require adaptability,
intelligence, and efficiency, necessitating a multidisciplinary approach that integrates
communication theory, machine learning, security, and hardware design.

\section{Conclusion}

This paper has offered a thorough analysis of channel coding approaches for 6G distributed
networks, showcasing their development, present proficiency, and relevance to next-generation
wireless systems. We started with a review of classical and modern coding techniques, the basics
of channel coding, and then moved on to more advanced codes, such as LDPC, Turbo, Polar, and
Fountain codes. Low-latency communication, enormous connectivity, energy efficiency, and
architectural decentralization are some of the fundamental requirements that we investigated in
relation to each code in 6G. Edge computing, unmanned aerial vehicle/satellite connection,
internet of things (IoT), and tactile internet are just a few of the many 6G applications that we
examined for performance across important use cases. Additionally, we assessed state-of-the-art
developments such as coding for cell-free massive MIMO, RIS-aware methods, AI-assisted and
quantum-enhanced coding, and more. Lastly, we highlighted the necessity for secure coding
structures that are flexible, intelligent, and able to adapt to different service and device needs
in real-time, and we listed important obstacles and future research objectives. Designing
next-generation IoT systems is profoundly affected by this study's consequences. Central to the
implementation of the Internet of Things are the examined channel coding methods, which bring
to light important compromises among dependability, delay, power usage, and decoding complexity.
When it comes to autonomous, decentralized Internet of Things (IoT) applications in fields like
smart cities, industrial IoT, and intelligent transportation, AI-assisted and quantum-enhanced
schemes provide new qualities of adaptability and resilience.

A cognitive architecture framework called CACADe-6G is also proposed in this article. It enables
real-time adaptive channel coding that aligns with the demands of dynamic 6G environments by
unifying sensing, learning, and control mechanisms. As detailed in Section~\ref{sec:cacade_eval},
this benefit manifests primarily as reduced switching overhead and energy savings, together with
robustness to correlated channel disturbances, rather than as a guaranteed BER reduction over
simpler memoryless adaptive baselines under i.i.d.\ perception noise. The analytical evaluation
presented there motivates three concrete next steps that we elaborate on below: full bit-level
Monte Carlo validation of the cognitive loop, testing under correlated and bursty SNR disturbances
such as intermittent jamming or shadowing, and hardware prototype validation of the CACADe-6G
architecture.

There will be a number of game-changing developments in 6G channel coding more broadly. Future
codes should be able to adapt and be aware of their context so that they can change things like
redundancy and rate in real time, depending on feedback from the network and the priorities of
services. The integration of learning-based coding and decoding frameworks with communication
stacks is expected to increase, leading to AI-native communication and the possibility of
performance and resource efficiency advantages. Maximum end-to-end latency and reliability can be
achieved through cross-layer optimization, which entails improving coordination between the
coding and other protocol layers, such as scheduling, routing, and modulation. For the long-term
viability and cost-effectiveness of supporting a network of trillions of connected devices,
energy-efficient designs are essential. Ensuring durable, efficient, and secure communication in
complex, distributed network systems will continue to be a key role of channel coding as 6G
evolves. For the Internet of Things (IoT) networks of the future, this research adds to the goal
of building AI communication agents that are smart, flexible, and reliable. In the future, we
would like to see this work used as a starting point for additional research in this exciting and
important field.



\bibliographystyle{elsarticle-num}   
\bibliography{sample}

\end{document}